\documentclass[pra,letterpaper,onecolumn,showpacs,superscriptaddress,floatfix]{revtex4} 
\usepackage{graphicx,psfrag,amsmath,amssymb,amsfonts,latexsym,color,dcolumn,bm,bbm} 

\begin{document} 

\title{Casimir forces and non-Newtonian gravitation}
\author{Roberto Onofrio} 
\affiliation{Department of Physics and Astronomy, Dartmouth
 College, 6127 Wilder Laboratory, Hanover, NH 03755, USA}
\affiliation{Dipartimento di Fisica
``Galileo Galilei'', Universit\`a di Padova, Via Marzolo 8, Padova
  35131, Italy}
\date{May 29, 2006}

\begin{abstract} 
The search for non-relativistic deviations from Newtonian gravitation
can lead to new phenomena signaling the unification of gravity with the 
other fundamental interactions. 
Various recent theoretical frameworks indicate a possible window for
non-Newtonian forces with gravitational coupling strength in the micrometer range. 
The major expected background in the same range is attributable to the
Casimir force or variants of it if dielectric materials, rather than 
conducting ones, are considered. 
Here we review the measurements of the Casimir force performed so far in the 
micrometer range and how they determine constraints on non-Newtonian
gravitation, also discussing the dominant sources of false signals.
We also propose a geometry-independent parametrization of all data in terms 
of the measurement of the constant $\hbar c$. Any Casimir force 
measurement should lead, once all corrections are taken into account,
to a determination of the constant $\hbar c$ which, in order to assess 
the accuracy of the measurement, can be compared with its more 
precise value known through microscopic measurements.
Although the last decade of experiments has resulted in solid 
demonstrations of the Casimir force, the situation is not 
conclusive with respect to being able to discover new physics. 
Future experiments and novel phenomenological analysis will be 
necessary to discover non-Newtonian forces or to push the window 
for their possible existence into regions of the parameter space 
which theoretically appear unnatural. 
\end{abstract}


\maketitle

\newpage

\section{Introduction}

The search for deviations of non-relativistic nature from Newton's universal 
gravitation has been a recurrent issue since the initial proposals dating 
back to more than thirty years ago \cite{Fujii,Fujii1,Fujii2,Zee,Zee1,Ellis}. 
In the last two decades, further motivation has been added by the possibility 
to observe deviations from the equivalence principle due to a force with 
coupling of order of the gravitational one, acting on a short range, 
and coupled to the baryonic number \cite{Fishbach1}.  
Although the initial claim for such a force, also named {\sl fifth force}, has 
not been confirmed by subsequent experiments, this has generated a revival 
of experiments in gravitation with a diversity of ingenious configurations. 
Thanks to these experiments, we now have a deeper knowledge of the 
gravitational field in the region from a few millimetres to planetary distances. 
The hypothetical forces superimposed on Newtonian gravity are typically parametrized 
by a Yukawa range $\lambda$ and a coupling strength $\alpha$ with respect to gravity, 
such that the total potential is 
$V_{\mathrm tot}(r)=V_{\mathrm N}(r)+V_{\mathrm Y}(r)$, 
where $V_{\mathrm N}(r)$ is the Newtonian potential and $V_{\mathrm Y}(r)$ the 
Yukawa potential. For a pointlike source of gravitational mass $M$, the two 
contributions respectively are:
 
\begin{equation}
V_{\mathrm N}(r)= ~- G  ~ \frac{M}{r}, ~ ~ ~ ~ 
V_{\mathrm Y}(r)= ~ -\alpha ~ G ~ \frac{M}{r} e^{-r/\lambda}.
\end{equation} 

\noindent
where $G$ is the Newtonian constant of gravitation. 
It is understandable that, due to the exponential suppression in the distance dependence 
do to a finite $\lambda$, even for a large $\alpha$ the signal contribution due to the Yukawian term 
will be masked by the long-range Newtonian term unless experiments are performed 
on a scale comparable or smaller than $\lambda$. 
Such Yukawian forces are expected in many attempts to unify gravity to
the other fundamental interactions. While variegate mechanisms have
been proposed so far predicting gravitational-strength Yukawian
forces \cite{Anto}, we would like to remark on the 
fact that even in minimal scenarios aimed at incorporating gravity into the 
standard model one should expect such forces.  
Indeed, by assuming that spontaneous symmetry breaking is the way to 
consider diverse interactions as different manifestations of a unified 
picture, the unification should give rise to the presence of new
intermediate gauge bosons with characteristics in between the parent interactions. 
The successful electroweak unification is a prototypical example, as
it predicts a new gauge boson, the $Z^0$ particle, with features in between the 
purely electromagnetic sector (it is electrically neutral like 
photons) and the purely weak sector (it is massive similarly 
to the $W^{\pm}$ bosons). By analogy, if we believe that  
spontaneous symmetry breaking is the universal path for unifying  
fundamental interactions (which will be scrutinized with the forthcoming  
experiments at the Large Hadron Collider), we should expect
short-range gauge bosons with intermediate coupling between usual gravity and 
the other interactions. 
Unfortunately, without further assumptions, the mechanism of
spontaneous symmetry breaking does not give quantitative hints on the couplings 
and range of the interactions. These parameters can be only determined from experimental 
inputs through new phenomena, which will be the counterpart of the
discovery of the neutral weak currents (and the consequent 
measurement of the Weinberg angle) for the successful electroweak unification.
\begin{figure}[t]
\begin{center}
\includegraphics[width=0.80\columnwidth,clip]{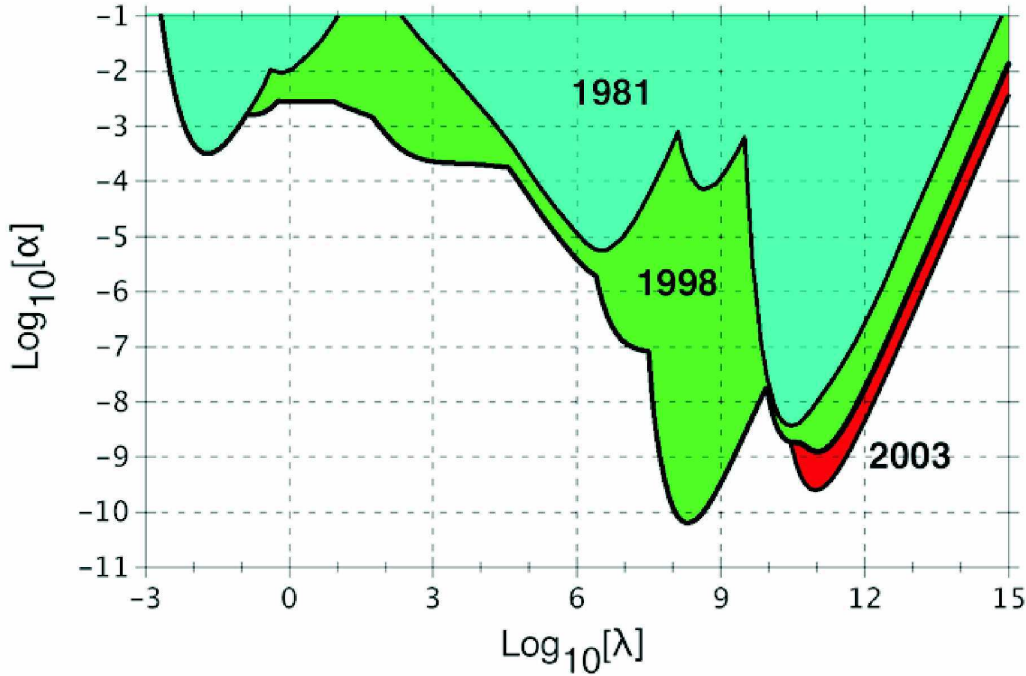}
\caption{Limits to the existence of new macroscopic forces parametrized 
by the coupling strength $\alpha$ (normalized to Newtonian gravity) 
and the Yukawa range $\lambda$ (in metres), coming from various experiments 
and astronomical observations. The progress made in the knowledge 
of the gravitational field in the macroscopic range since the 
proposal for a ``fifth force'' has been considerable.
Limits are however rather weak at the two extreme lengthscales, 
in particular at the interesting submillimetre region for $\lambda$ 
from the viewpoint of extra-dimensional models (courtesy of  
Coy, Fishbach, Hellings, Standish, and Talmadge, 2003).}
\end{center}
\label{fig1}
\end{figure}
This motivates the search for non-Newtonian forces in the broadest 
range of distances and couplings. The current limits in the 
$\alpha-\lambda$ plane, with $\lambda$ spanning from 1 mm 
to $10^{15}$ m, are shown in Fig. 1. It is recognizable that the 
constraints on extra-gravitational, short-range forces in the 
submillimetre range, and in the range above $10^{15}$ m, 
do not match the more stringent limits in the intermediate region.   
On the other hand, various models based on supersymmetry predict the
existence of forces superimposed to gravity in the range between 
1 $\mu$m and 100 $\mu$m with intensity up to $10^5$ times the 
strength of gravity in the same range \cite{Antoniadis,Arkani1,Arkani2}. 
The corresponding limits to their existence in this range are quite weak.  
Although in principle a measurement performed at any lengthscale 
(for instance a high-accuracy measurement of the gravitational 
force in the metre range with a Cavendish experiment) may constrain 
values of $\lambda$ at very different lengthscales, the exponential sensitivity of 
$\alpha$ to $\lambda$ in the limit $\lambda \rightarrow 0$ gives rise
to very weak bounds for $\lambda$ smaller than the lengthscale 
at which the experiment is performed. 
Therefore, in order to give significant limits in the micrometre
range, one may opt for designing experiments intrinsically exploring small distances. 
In doing so a signal-to-noise issue is soon faced since, as 
first discussed in \cite{Price}, the Newtonian gravitational force 
signal scales as the fourth power of the size of the experiment, with 
$F_N \propto M^2/r^2 \simeq \rho^2 r^4$, where $M$ and $\rho$ are mass
and density of the test bodies and $r$ their distance. This implies that background 
originating from a force having a slower scaling with distance is
going to strongly affect our capability to obtain significant limits.  
In the range of $\simeq$ 100 $\mu$m, limits obtained using Cavendish
balances \cite{Hoskins} have been recently improved by the group at 
the University of Washington using torsion pendula and rotating attractors
\cite{Smith,Hoyle}.  Decreasing the explored range of distances below
the 100 $\mu$m range is challenging due to the requirements necessary 
to ensure parallelism between the two surfaces during the rotation.  
An alternative technique to explore the range below 100 $\mu$m consists
in the use of microresonators and nanoresonators combined to 
dynamical detection techniques first introduced in 
\cite{Price,OnofrioCarugno,CarugnoOnofrio}.   
Two groups, at Stanford and Boulder, have performed measurements based
on this technique and limits in the
$\alpha-\lambda$ plane have been reported \cite{Chiaverini,Long}. 
The presence of an electromagnetic shield between the source of the 
non-Newtonian field and the microresonator used for the detection 
of the force limits the minimum explorable distance.

On the other hand, the most promising theoretical predictions appear
to be in the range of 1-10 $\mu$m, and therefore it is important to
design experiments which maximize the chance of detecting forces 
acting in this distance range. In order to minimize the background, and
taking into account the huge difference in coupling strength between 
gravitation and electromagnetism, it seems compulsory to consider 
electrically neutral bodies. Then, the strongest expected background 
in the micrometre range for forces between electrically neutral 
bodies is due to the Casimir force \cite{Casimir}. 
This also means that any accurate measurement of Casimir forces 
gives limits on gravitational-like forces, as remarked 
in pioneering papers by Mostepanenko and Sokolov 
\cite{MostepanenkoSok1,MostepanenkoSok2} two
decades ago. Since then,  precision studies of Casimir forces in the 200 nm-2
$\mu$m range  have allowed us to grasp a better understanding of the
background in which novel macroscopic forces of gravitational origin
could be immersed. 

Many excellent reviews on various aspects of 
Casimir physics have been available since two decades 
\cite{Plunien,Elizalde,MostepanenkoTrunov1,Milonni,MostepanenkoTrunov2,Bordag,Bordag1,Reynaud,Milton,Milton1,Lamoreauxrev}, 
and more recently reviews have also appeared on the experimental efforts 
to detect non-Newtonian gravitational forces above the range of 
10-100 micrometres \cite{Fischbach,Long2,Long3,Adelberger}. 
Considering the abundance of existing review material on both Casimir forces and 
the search for non-Newtonian forces, we will only cover some 
selected topics lying between the two groups of reviews \cite{Gundlach}. 
In particular, we will discuss limits to non-Newtonian 
forces in the micrometre range in which Casimir forces are expected 
to be the leading background. Obviously, this review should be considered 
as a progress report since the situation is evolving 
quickly due to the expected flow of new data and more stringent 
phenomenological and theoretical analysis.  
The spirit will be also to keep the formalism to a minimum  and to
put more emphasis on physical and dimensional grounds, to allow as
much as possible for a pedagogical, broad-audience approach. 
After a discussion of the Casimir force and its meaning and importance 
in current physics, we discuss its experimental evidence and comment 
on the consequent limits to gravitational forces. 
We then discuss the variety of leading systematic effects in the
experiment, and propose a parametrization of the Casimir experiments which 
allows for a general comparison between various configurations.
Some promising future research directions are then discussed, with more 
general remarks in the concluding section.

\section{Casimir forces and quantum vacuum}

The years 1947 and 1948 were milestones for the development of quantum field 
theories since almost simultaneously Lamb shifts in hydrogen were measured 
in high precision spectroscopy experiments, renormalization was proposed 
by Feynman, Schwinger, and Tomonaga, and a macroscopic force originating from 
quantum vacuum fluctuations was predicted by Casimir.
The following six decades of developments in microscopic physics 
may be considered as the refinement of these findings in quantum 
electrodynamics (QED) and its extension to other fundamental interactions. 

With regard to QED in the macroscopic realm, one may ask if it 
leads to macroscopically detectable forces without classical counterpart. 
Evidently, if quantum vacuum fluctuations are isotropic, no force should be expected 
in the simplest case of an infinite mirror as the contributions 
due to the virtual photon radiation pressure will be equal on both
sides. Such an isotropy is instead broken considering two indefinite 
and parallel plates. A simple stationary boundary condition for the 
electromagnetic fields on the surfaces, for instance the one typical 
of an ideal conductor, gives a drastic change in the density of states 
of the virtual photons inside the cavity with respect to that 
outside the cavity. When the net radiation pressure is evaluated, there 
is now a partial cancellation leading to the expression of the attractive \cite{Note1}
Casimir force per unit of surface area $P_C$ \cite{Casimir}

\begin{equation}
P_C=\frac{K_c}{d^4},
\end{equation}

\noindent
where $d$ is the distance between the two plates and $K_c$ is a constant. 
We want to stress some qualitative features of this very simple formula.

\vspace{2.0mm}  
\noindent
{\sl i) The Casimir pressure depends on the product $\hbar c$ and 
does not depend on the electric charge at the leading order.}
The concrete expression for $K_c$ contains only the two constants 
of nature allowed in free quantum field theory, {\it i.e.} 
$\hbar$ and $c$, being $K_c= \pi^2\hbar c/240$. 
Indeed, all formulae for the Casimir force considered here 
have in common the dependence on the product $\hbar c$. 
In a classical, even if relativistic, setting (no quantum fluctuations, 
{\it i.e.} $\hbar \rightarrow 0$) or in a non relativistic, even if quantum, realm 
(by hypothetically imagining an infinite speed of light, $c^{-1} \rightarrow 0$, 
in  which case however electromagnetism would be quite different) there are 
no Casimir forces. This is a distinctive feature with respect to the 
van der Waals-London forces in which relativity and the consequent 
finite speed of light propagation do not play any direct role. 
In fact, the Casimir force was initially originated from the attempt 
to include the relativistic retardation into the London theory. 
We will exploit this feature of the Casimir force later on in 
discussing a comparison among different geometries. 
Also, although we are considering electromagnetic fluctuations, the electric
charges which mediate the mechanical pressure between the virtual
photons and the bulk material of the mirror do not appear explicitly. 
This apparently paradoxical property of Casimir pressure is also known 
as {\sl saturation} \cite{Reynaud,Jaffe}. The combined effects of the 
value of the electric charge and the reflectivity of the mirrors are 
large enough to make the force asymptotically independent of the same 
electric properties producing the force. 
This saturation property is expected to fail if the reflectivity of 
the mirrors is not complete, which happens for any realistic 
material for electromagnetic frequencies above a finite cutoff frequency. 
In this case there is a next-to-leading correction to the Casimir 
force depending upon the electric charge and the finite conductivity 
properties of the material.  
Apart from the electromagnetic field, other fundamental interactions 
do not give significant contributions to the Casimir pressure. 
For gravitational quantum fluctuations, apart from the impossibility to
identify a gravitational conductor, the force is too weak to play a 
role and cannot give a saturated contribution. 
Strong and weak fields mediated by gluons and $W^{\pm}, Z^0$ 
particles cannot play a role in the micrometre range 
since their influence is limited to the subatomic domain. 

\vspace{2.0mm}
\noindent
{\sl ii) The Casimir force may be interpreted as due to the 
interplay between quantum fluctuations and geometry.}
The Casimir force has been interpreted in a very general way 
as emerging from the finite cancellation between the radiation 
pressure due to virtual photons present in quantum vacuum and 
acting on a macroscopic body. Seen from this perspective, the Casimir 
force is a strongly geometry-dependent force and is a macroscopic 
manifestation of the effect of boundary conditions on quantum fields,
a manifestation of a sort of {\sl quantum electrostatics}. 
In addition, it provides physical insights into the same concept of 
renormalization in quantum field theory. In principle, the 
finite cancellation of radiation pressure originates from the 
difference between diverging contributions, although in practice the 
high-frequency response of the conductors provides a physical regulator. 
Alternative views centered on the role of the sources of the
electromagnetic field have been also proposed, based on the idea that 
the geometric, field theoretic interpretation of the Casimir result 
is more an issue of taste rather than an essential, substantial
physical point \cite{Milonni1,Milonni2,Milonni3} and that quantum 
fluctuations are actually originating from the sources of the fields. 
This is key in the calculation of the same effect by Lishitz
\cite{Lifshitz}, using the fluctuations of the electric and magnetic 
fields of the sources in the materials.  
Therefore there are two distinct approaches to the same effect 
described by Eq. 2, the one adopted by Casimir, more idealistic, 
based on global, geometrical properties of quantum fluctuations, and 
that by  Lifshits, more materialistic, based on local, analytical 
properties of the sources. 
This dualism in the interpretation of the Casimir force still exists today, 
and the relative contributions of intrinsic field quantization and non-linear 
features of source models have been discussed 
(see for instance \cite{Schwinger1,Schwinger2,Milonni4,Milonni5,Cohen1,Cohen2}). 
The issue is currently important at the cosmological level, as the actual 
presence of quantum fluctuations at the macroscopic level gives rise 
to the outstanding problem of their expected large contribution to the
cosmological constant \cite{Reynaud,Zeldovich,Weinberg,Peebles}. 
The presence of this contribution is instead an ill posed problem if one 
attributes no reality at all to intrinsic quantum field fluctuations, as 
discussed in \cite{Jaffe}. 

We remark on the fact that a force proportional to $\hbar c$ seems 
to naturally support a field theoretic interpretation in terms of irreducible 
quantum fluctuations. In fact, in the quantum vacuum we pictorially expect 
a particle-antiparticle pair to appear and disappear on a timescale 
$\tau \simeq \hbar/m c^2$ for massive particles (characterized by an 
energy-momentum dispersion law $E^2=p^2 c^2+m^2 c^4$), and 
$\tau \simeq 1/\omega$ for massless particles characterized by 
an energy-momentum dispersion relationship $E=p c=\hbar \omega$. 
The maximum distance the virtual particles may travel to preserve 
its virtuality cannot exceed the quantity $c \tau$. Then the product of the 
virtual energy transported in space times the traveled distance is limited 
to values smaller than $E c \tau \simeq \hbar c$, {\sl regardless} of the particle mass. 
From this point of view, if we think of the Casimir force as originated from 
quantum vacuum fluctuations, it is not surprising that the only fundamental 
quantity appearing is then the product $\hbar c$, for any virtual particle. 

Along these qualitative lines, we can in fact  understand the Casimir formula by 
using dimensional arguments. In a parallel plane situation there is only a finite 
length available, {\it i.e.} the distance between the two plates $d$. 
A pressure is dimensionally equivalent to an energy density, and we can form an 
energy from $\hbar c$ (energy times length) considering $\hbar c/d$. 
An energy density is obtained by further dividing by $d^3$, leading to 

$$P_C \propto 
\frac{\mathrm{Energy} \times {\mathrm{Length/Distance}}}{\mathrm{Volume}} 
\propto \frac{\hbar c/d}{d^3},
$$

\noindent
which gives the proper scaling with the fourth power of the distance. 
This elementary dimensional argument is of less unambiguous application 
in other geometries since more than one finite length plays a role, for 
instance the radius of the sphere and the distance between sphere and 
plane in the case of a sphere-plane geometry.
  
The study of the Casimir force is becoming an interdisciplinary 
subtopic of physics whose importance is growing in many fields. 
In cosmology and astrophysics, quantum vacuum is a candidate to 
support the {\sl dark energy} hypothesis 
\cite{Peebles,Weinberg1,Sahni,Witten,Carroll,Genet,Caldwell} which is 
considered responsible for the observed acceleration of the Universe
\cite{Riess,Garnavich,Perlmutter}.
In nanotechnology these forces may play a significant role in 
many artificial structures built on the nano-to-microscale \cite{Serry}. 
Even in apparently unrelated fields, such as biophysics and chemistry, 
many of the relevant adhesion forces may call for a more complete  
framework than the one available through the van der Waals force, 
of which the Casimir/Casimir-Polder forces appear as the relativistic 
generalization \cite{Parsegian}. 
This interdisciplinary flavor has grown since what was considered
a scientific curiosity (a minute force between macroscopic and 
electrically neutral surfaces) has been increasingly explored in 
the laboratory. The laboratory findings have been complemented by 
theoretical insights then converted into further experimental 
investigations, in a close relationship between theory and experiment. 

\section{Experiments on Casimir forces}

After the prediction of Casimir, a first generation of experiments was 
performed in the following decade. Since a detailed description of these 
experiments and a critical assessment of their limitations have 
been covered in a recent review \cite{Bordag1} we will limit 
attention to brief considerations of the successful experiments 
with particular regard to some general methodological comments. 
It was soon recognized that the difficulty in the verification 
of the Casimir formula was not the expected force signal, which 
is relatively large at distances of order of few micrometres 
for macroscopic (at least few mm$^2$) surface areas, but in the
achievement and assessment of the parallelism, as well as in 
precisely measuring the distance between the plates. 
An experiment in the original conducting parallel plate 
situation studied by Casimir was performed by Sparnaay
\cite{Sparnaay}, while other experiments were performed on
configurations different either in the geometry or in the 
conductivity properties. The sphere-plane geometry was investigated 
by van Blokland and Overbeek \cite{vanBlokland} to overcome the
difficulty of maintaining the two plates parallel within the 
required accuracy to test the Casimir formula. 
The use of dielectric surfaces was proposed to have an 
accurate assessment of the gap distance, for instance by optical 
techniques. Obviously both these experimental shortcuts have 
associated issues with regard to the experiment-theory comparison. 
In the case of the sphere-plane geometry, an exact evaluation 
has only been obtained for a scalar field very recently \cite{Bulgac},
and all the experiments had to rely on an approximation 
used in classical electrostatics, known as Proximity 
Force Approximation (PFA) \cite{Derjaguin,Derjaguin1,Blocki}. 
This approximation gives reliable results if the relevant surfaces are 
separated by a distance much smaller than their typical local curvatures. 
This leads to an approximate expression for the Casimir force

\begin{equation}
F_C(d)= \frac{\pi^3 \hbar c}{360} \frac{R}{d^3},
\end{equation} 

\noindent
where $R$ is the radius of the sphere and $d$ is its distance 
from the plane, and its validity holds in the regime $R>>d$. 
Since this approximation can be derived in classical electrostatics 
by relying on the additivity of the Coulomb force, care has to be
taken in the case of forces of quantum nature which have a strong 
geometric (and non-additive) character. 
Likewise, in the case of dielectric materials the comparison 
with theory is complicated by the necessity to know the dispersive properties 
of the dielectric material. This is taken into account with a formula 
developed by Lifshitz and collaborators \cite{Lifshitz,Lifshitz1}. 
The loss of {\sl universality} intrinsic in the ideal Casimir formula 
makes the Lifshitz formula less appealing and complicates the 
theory-experiment comparison requiring a detailed knowledge of 
the dielectric response of the materials. 
Nevertheless, the problem was not felt to be important, as the experimental
precision was limited in comparison to the more
stringent tests of quantum electroynamics at the microscopic level with Lamb shifts and $g-2$ 
for electron and muon; no need for refined comparison with theory 
was then necessary. 

The outcomes of the first generation of measurements can
be summarized as follows. The Sparnaay experiment, with accuracy 
assessed at the 100$\%$ level, was considered as inconclusive in 
showing the expected scaling of the force with the distance, with also  
evidence for repulsive forces indicating a partial control over the 
electrostatic background. To use Sparnaay's own words, the measurement 
``{\sl did not contradict Casimir's theoretical prediction}.'' 
The experiment by van Blokland and Overbeek was more successful from 
this viewpoint, obtaining agreement with the Casimir predictions 
at an estimated accuracy around 50$\%$, and was thus the first
uncontroversial verification of the Casimir force between 
metallic surfaces. Experiments with dielectric surfaces were 
performed using silica lenses \cite{Derjaguin,Abrikosova}, 
crossed cylinders of muscovite mica \cite{Tabor,Israelachvili}, thin films 
of liquid helium absorbed on surfaces of alkaline-earth fluoride 
crystals \cite{Sabisky}, flat surfaces of porosilicate glass 
\cite{Kitchener}. The evidence for a crossover from the non-retarded component 
of the molecular force to the retarded component and an overall 
verification of the Lifshitz theory at the 20-40 $\%$ accuracy level, apart from the 
experiment by Sabinsky and Anderson reporting accuracy of order 1$\%$,  
were the main results of these experiments.  

After this burst of experimental activity on Casimir forces there 
was no further activity for many years. The spectacular success 
of quantum electrodynamics and its unprecedented accurate
verifications at the microscopic level could not be matched by
measurements of (necessarily macroscopic) forces. 
The attention at the macroscopic level was instead shifted on 
the atomic physics experiments, as the presence of a cavity with 
defined boundary conditions was found to affect the spontaneous 
emission properties of individual atoms. 
In this context, the microscopic counterpart of the Casimir force 
acting between an atom and a plane surface, also known as Casimir-Polder 
force \cite{Polder}, was measured by looking at the deflection induced 
on an atomic beam by two parallel plates \cite{Sukenik} and comparing
this with the theoretical predictions \cite{Bartonpolder}. 
The new wave of Casimir force experiments 
was revamped after remarks by Sparnaay \cite{Sparnaaymaking} 
concerning the possibility of  a second generation of measurements
at higher accuracy exploiting the emerging subfields of 
atomic force microscopy (AFM) \cite{Binnig} and of nanotechnology. 
Consequently, in partership with Carugno at the INFN in Padova, 
we studied a scheme to measure the Casimir force in a parallel plate 
configuration, starting the first tests in the early summer 1993 \cite{noteono}. 
The apparatus capitalized on a variety of technological improvements not
available at the time of the Sparnaay's measurement.
 Most notably, the use of microresonators and of 
dynamical detection techniques based on the Fourier analysis of the 
tunnelling current of a single axis scanning tunnelling microscope 
were discussed and a first prototype tested inside a scanning electron 
microscope.  Also, consideration was given to the capability of 
measuring the gravitational force in the same range \cite{OnofrioCarugno}.
Unfortunately the issues of parallelization, dust in the gap, and 
the large $1/f$ noise present in electron tunnelling devices prevented 
a straightforward measurement of the Casimir force in the proposed 
configuration. 

An attempt to measure the Casimir force using a torsional balance 
was initiated by Lamoreaux at the University of Washington in Seattle. 
The initial tests with flat plates, in 1994, also met difficulties 
in the alignment \cite{Lamoreauxhisto}, until the experiment was 
reconfigured in the sphere-plane geometry by using a convex lens, and 
the Casimir force was then measured at distances up to 6 $\mu$m
\cite{Lamoreaux} with significant improvement in both range and 
accuracy with respect to the van Bockland and Overbeek measurement 
in the same configuration. These improvements were mainly due to the 
elimination of mechanical hysteresis in the torsion balance and the 
use of piezoelectric actuators for the positioning of the plates \cite{notelam}.

Due to the large Volta potential present between the plates even after 
a nominal external short-circuit, even at the closest explored
distance the Casimir force was evaluated to be about 20 $\%$ of the
total measured force, and required an ingenious subtraction technique to be employed.  
Theoretical discussions followed the appearance of the related paper, 
focusing on finite conductivity and temperature corrections. 
Given the large range investigated, this experiment with the accuracy 
initially quoted was in principle able to grasp both these corrections. 
A deeper analysis showed that the conductivity corrections were less trivial 
to manage due to the presence of a copper substrate deposited on the 
lens prior to the gold coating. Including a better assessment of the radius 
of curvature of the lens, found {\it a posteriori} to be aspheric
\cite{Lamoreaux1}, did not solve the conductivity issue. 
Further discussions of the experiment were carried out regarding the 
conductivity corrections \cite{LambrechtLamo,LamoLambrecht}
 and the thermal corrections \cite{Bostrom,LamoBostrom,SerneliusLamo}.
While we suggest that the reader looks at the related interesting
exchange of comments, a likely assessment of the situation can be summarized 
as follows: the initially quoted accuracy of 5$\%$ was probably 
reliable at the smallest explored distances, but it was worse at the 
largest distances. Lamoreaux himself pointed out the spirit of his 
measurements in one of the abovementioned replies \cite{LamoLambrecht}: 
``{\sl I offer the caveat that my experiment was intended 
as a demonstration to show that, with modern experimental 
techniques, one could do a really accurate measurement 
of the Casimir force. As a demonstration, only minimal 
tests for possible systematic errors were performed: 
furthermore, I was satisfied with the agreement between 
my experimental result and my inaccurate calculation}.'' 
This remark by the pioneer of the modern generation 
of measurements on Casimir forces, as we will see in the following
sections, is key for understanding the spirit with which the current generation 
of measurements on Casimir force has been carried out: they have 
to be considered more as demonstrations than experiments \cite{notedem}.

The successful use of atomic force microscopy techniques combined 
with the sphere-plane geometry was accomplished by Mohideen and Roy 
at Riverside in 1998 \cite{Mohideen1}, after attempts started 
one year earlier. In their experiment, a metallized polysterene
sphere was mounted on the tip of the AFM cantilever, and the deflection of 
the cantilever measured as a function of the distance between the 
sphere and a metallized flat surface. The metal deposited on the 
sphere was initially aluminum but  a second version of the 
experiment instead used gold \cite{Mohideen2} which was predicted 
to provide a cleaner situation \cite{Svet}. 
In both cases, the experiment-theory comparison took a number of 
corrections into account, namely the finite conductivity, 
the roughness, and the finite temperature, unobserved in the Lamoreaux 
measurement. Due to the small range of distances investigated, down to
100 nm, and the smaller Volta potential, of order  30 mV, the Casimir
force dominated the electrostatic contribution over a wide 
range of distances, with the latter contributing to the bare force 
only in an amount evaluated as less than 3$\%$. 

\begin{figure}[t]
\begin{center}
\includegraphics[width=1.0\columnwidth,clip]{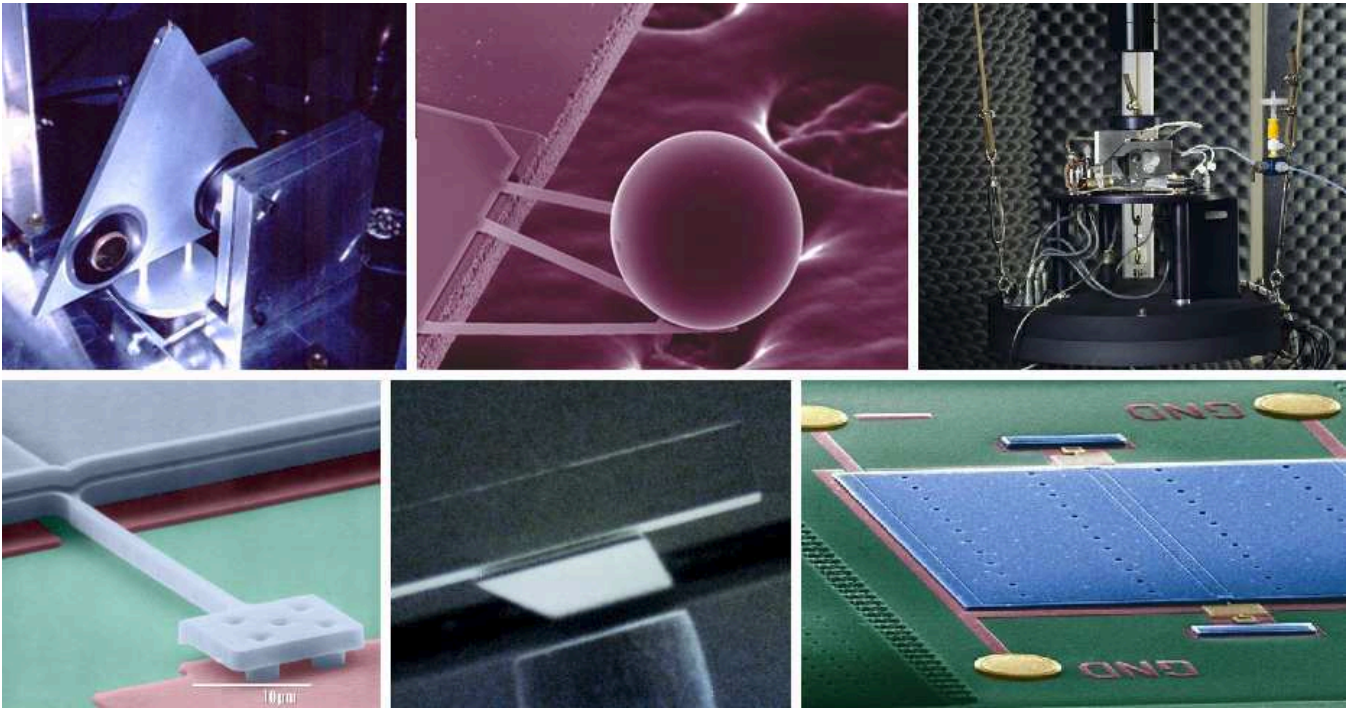}
\caption{Pictures from the six Casimir force experiments of the second
  generation. In chronological order, from top-left to bottom-right, 
some views of the apparatuses used in Seattle, Riverside, Stockholm, 
Murray Hill, Padova, and Indianapolis are depicted.}
\end{center}
\end{figure}

A third successful attempt was performed by Ederth \cite{Ederth}. 
This work is remarkable for a number of reasons. The force 
was measured in the novel geometry of crossed cylinders (previously 
only used by Tabor and Winterton \cite{Tabor} for dielectric surfaces), and 
in the very short range of 20-100 nm. The expected Casimir force in 
such a configuration can be written as:

\begin{equation}
F_C(d)= \frac{\pi^2 \hbar c}{360} \frac{{(R_1 R_2)}^{1/2}}{d^3}
\end{equation} 

\noindent
where $R_1$ and $R_2$ are the radii of curvature of the two cylinders.
Also, extreme care was put into polishing and chemically treating
the surfaces, as well as assessing their roughness. Finally, a critical 
reanalysis of the precision was performed both with reference to the 
acquired data, and with respect to previous experiments in which 
the experimental precision was obtained with reference to 
theoretical predictions, in the spirit of a demonstration 
of an {\it a priori} known law. 
A further issue raised from Ederth was that the Casimir force 
is of such a nonlinear nature and spanning such a large range of
values that it makes it difficult for an analysis with the rms error
approach. He also noticed that an average, arbitrarily extended over a large region in 
which the force is small, will artificially decrease the rms error 
(see Fig. 6 in \cite{Ederth}). Moreover, the author discussed 
the effect of an unwanted oversight on the determination of the 
absolute distance between the two test bodies, concluding that 
if the separation between the surfaces is not measured directly 
with high precision, the induced uncertainty will make the 
relative error large (see Fig. 7 in \cite{Ederth}). 
While these may appear as details in the framework of demonstrating 
the Casimir force, they are actually critical if one wants to 
assess limits to new physics.  

One variant of the sphere-plate measurement was performed at Bell 
Laboratories by a team led by Capasso \cite{Chan1}. 
The force was measured between a sphere (similarly to the Mohideen set-up) 
but using a micro-electromechanical (MEM) resonator as the flat
surface. The torque exterted by the sphere on the resonator 
unbalanced the capacitive bridge which used a second set of 
plates under the MEM. The rms deviation between theory and 
experiment was less than 0.5 $\%$, but the authors were more 
conservative and considered this quite accidental in the light 
of the presence of various factors not controllable at the same 
level of accuracy, most notably the spread in the optical properties 
of gold films. The main goal of this measurement was to show that 
Casimir forces are of relevance in micromechanics and (at least 
until they do not correspond to non-retarded components) in 
nanomechanics, and the interesting nonlinear dynamics predicted in 
\cite{Serry} including Casimir switching was also demonstrated \cite{Chan2}. 
This experimental effort continues at Harvard, focusing 
on the influence on Casimir forces of materials with modulable 
reflectivity properties  \cite{Iannuzzi1}, torque induced by vacuum 
fluctuations \cite{Iannuzzi2}, and effect of the finite thickness 
of the metallic deposit \cite{Iannuzzi3} (see also \cite{Iannuzzi4,Iannuzzi5}
for reviews of these efforts).
Previous claims of a possible role of the Casimir force in MEMs can 
also be found in \cite{Buks1,Buks2}, but the related experiments found
a discrepancy of a factor six with respect to the Casimir prediction. 
It has been suggested by the same authors that this discrepancy arose   
from source roughness or an adsorbed layer between the two surfaces,
but regrettably no further investigations were carried out to solve this puzzle. 
An alternative and more conservative explanation was suggested in the 
conclusions of \cite{Chan2} as due to the electrostatic forces
estimated to be $10^6$ times larger than the expected Casimir force.

The initial attempts started in Padova to perform a measurement in the parallel 
plane configuration became successful in the late Spring 2001 \cite{Bressi}. 
The key upgrades with respect to earlier efforts, as described in
detail in \cite{Bressi0,Bressi1}, were the use of a 
fiber optic interoferometre as a displacement detector \cite{Rugar}, 
the monitoring and cleaning of the two surfaces from dust with an 
{\it in situ} tool, the implementation of frequency-shift techniques. 
The frequency-shift technique was already well-known in the field of 
Atomic Force Microscopy \cite{Giessbl} and its use in Casimir force experiments 
was first reported in \cite{Puppo,Bressi0,Chan2}. 
The measurement was performed in the $0.6-3 \mu$m range, with the 
lower limit approaching the expected Casimir stiction between the two 
surfaces considering the stiffness of the cantilever, and the upper 
limit due to the sensitivity of the fiber optic interfemetre. 
The data were analyzed using the plain Casimir formula without 
corrections, and the accuracy in the measurement of the Casimir 
coefficient (which can be directly translated into a measurement of 
$\hbar c$) was 15 $\%$. Major limitations to the accuracy were the 
precision in determining the absolute distance from the electrostatic
calibration, the finite amount of parallelism, and the piezoelectric 
actuator with 10$\%$ deviation from the expected specifics, which 
was overcome with independent calibrations.  
Attempts to introduce the finite conductivity in the data fit led to 
better accuracy estimated to be around 5$\%$ \cite{Mostepnote}. 
However, in a conservative approach, no effort was made on a more 
sophisticated data analysis mainly since the better fitting 
involved points at the smallest distances which were taken close 
to the snapping critical point (estimated, given the stiffness 
of the resonator, to be around 0.4 $\mu$m) potentially inducing 
anharmonicity and instability in the resonator motion. Furthermore, 
the experiment was felt to be  less accurate than previous sphere-plane 
experiments. Nevertheless, the result of this experiment represents 
a significant improvement over the pioneering but ambiguous result 
obtained by Sparnaay in the original plane-plane geometry studied by Casimir.
Further motivations to study this geometry have been discussed in \cite{OnofrioOK}.

The most recent reports on Casimir force measurements are from Decca
and collaborators at Indiana University. 
The Casimir force was initially measured in a configuration similar to 
the one studied at Bell Labs but using different metals \cite{Decca1,Decca2}. 
Apart from the use of different metals, gold and copper, the
experiment makes use of a fiber optic interferometre to maintain the 
fiber-platform separation constant. 
The same dynamical frequency-shift technique previously used in
\cite{Puppo,Chan2,Bressi} was implemented. The authors also claimed 
that the dynamical technique was equivalent to determining the Casimir 
pressure between two parallel plates, but this formal equivalency 
relies upon the validity of the proximity force approximation,
similarly to all previous sphere-plane measurements, and therefore 
this is a way to restate the same physics. 
This group has made a well focused effort to constrain non-Newtonian 
forces using a variety of schemes, with particular regard to the 
cancellation of the Casimir force using the so-called isoelectronic 
technique \cite{Fischbachiso}. 
In this approach, the differential force between a gold-coated sphere 
and flat surfaces of gold and germanium coated with a shared layer of
gold was measured, leading to a fractional difference in the Casimir 
force estimated to be $\simeq 10^{-6}$, below the experimental
sensitivity \cite{Decca3}. The presence of a force signal should be
due to a non-Casimir like interaction. A non-zero force was evidenced, 
but it was attributed to the residual Casimir force due to the 
different height difference between the two flat surfaces, not 
controllable to better than $\simeq 0.1$ nm. The experiment is planned 
to run using different isotopes of nickel to further reduce the 
fractional difference in the electronic properties. 
The more recent results have been questioned with regard to the 
claimed precision \cite{Decca4} in some of the parameters involved in the data 
analysis \cite{Miltonrecent}. This issue seems still controversial 
and in the next section we will discuss some tests which could be 
possible {\sl smoking guns} in assessing the sensitivity and the 
precision of all the Casimir force apparatuses. 

\begin{table}[t]
\begin{center}
\begin{tabular}{|l|c|c|c|l|}
\hline\hline
Year & Geometry          & Range ($\mu$m)   & Accuracy ($\%$)&  Reference 

\\ \hline  
1958 & plane-plane       & 0.3 $\div$ 2.5   & 100   & Sparnaay \cite{Sparnaay} \\ 
1978 & plane-sphere      & 0.13 $\div$ 0.67 & 25    & van Blokland \cite{vanBlokland} \\
1997 & plane-sphere      & 0.6 $\div$ 12.3  & 5     & Lamoreaux \cite{Lamoreaux} \\ 
1998 & plane-sphere      & 0.1 $\div$ 0.9   & 1     & Mohideen \cite{Mohideen1} \\ 
2000 & crossed cylinders & 0.02 $\div$ 0.1  & 1     & Ederth \cite{Ederth} \\ 
2001 & plane-sphere      & 0.08 $\div$ 1.0  & 1     & Chan \cite{Chan1}\\
2002 & plane-plane       & 0.5 $\div$ 3.0   & 15    & Bressi \cite{Bressi} \\ 
2003 & plane-sphere      & 0.2 $\div$ 2.0   & 1     & Decca \cite{Decca1} \\ 
\hline\hline
\end{tabular}
\end{center}
\caption{Status of the experimental studies of Casimir forces between 
metallic surfaces. For each experiment, listed in chronological order,
the investigated geometry, the explored range of distances, the
claimed accuracy, and the first author and reference are reported.
The claimed accuracy is the one quoted by each group, however it
often corresponds to different definitions with different statistical 
meanings. In the experiment by Sparnaay repulsive forces were observed between 
aluminum surfaces, while the quoted accuracies for the van Blokland and 
the Lamoreaux experiments have to be considered as reliable only at the 
smallest distances. Whenever there was a sequel of experiments by the
same group, the quoted reference is for the first reported data set.}
\end{table}

We summarize in Table I the current knowledge of Casimir forces between
metallic surfaces, while in Fig. 2 pictures from the six recent experiments are shown. 
In Fig. 3 the corresponding reported limits in the $\alpha-\lambda$
plane are presented. 

This succinct overview of the recent experiments on the Casimir forces 
needs to be complemented by the parallel developments in the measurement of the
Casimir-Polder force, the long distance forces acting between 
atoms and macroscopic surfaces \cite{Polder}. 
Casimir-Polder forces establish the link between the macroscopic Casimir forces and 
the microscopic atom-atom interactions (like the Van der Waals forces).
This force has been measured by looking at the deflection induced on an atomic beam of sodium 
atoms by the presence of a metallic surface \cite{Sukenik}. The
accuracy of this measurement was around 12 \%  limited by the control of the distance between 
the atomic beam and the surface, but it was enough to confirm that 
the Casimir-Polder force, and not the Van der Waals force, was 
acting in the configuration investigated. 
More detailed studies of Casimir-Polder force are necessary for 
a variety of reasons. It has been recently conjectured that
their study at larger distance can more easily provide information on the 
thermal contribution to the force \cite{Antezza1}. Interesting studies 
on non-equilibrium statistical mechanics are also possible \cite{Antezza2}, and 
their role in the formation of weakly-bound macromolecular states 
of biological interest has yet to be explored.   
The technological interest in these studies is motivated by the manipulation 
of cold atoms on chips for the possible implementation of 
quantum information schemes or for characterization of surfaces. 
Quantum operations can be performed only by preserving quantum
coherence, so it is important to evaluate the effect of surfaces 
in close proximity of cold atoms \cite{Fortagh,Leanhardt,Jones,Harber1,McGuirk,Lin}.
In a recent experiment, the Boulder group at JILA led by Cornell
succeeded in measuring the Casimir-Polder force \cite{Harber2}.
The dipole oscillation frequency of an elongated, cigar-shaped, 
condensate of rubidium atoms was determined versus the distance between the condensate 
and a conducting surface. This experiment could also soon lead to the 
observation of the thermal corrections to the Casimir-Polder force. 
In this case, the information could be used to infer the thermal 
corrections to the purely Casimir force, especially if the
investigations will be extended to dielectric surfaces. 
Another potentially fruitful technique is the use of atomic 
interferometry, with the first measurement of the van der Waals force 
reported in \cite{Cronin}.

\begin{figure}[t]
\begin{center}
\psfrag{x}[][]{\large ${\mathrm{Log}}_{10} [\lambda]$}
\psfrag{y}[][]{\large ${\mathrm{Log}}_{10} [\alpha]$}
\psfrag{a}[][]{Riverside}
\psfrag{b}[][]{Indiana}
\psfrag{c}[][]{Seattle}
\psfrag{d}[][]{Stanford}
\psfrag{e}[][]{Boulder}
\psfrag{f}[][]{E\"ot-Wash}
\psfrag{exclude}[][]{\large Excluded region}

\includegraphics[width=0.80\columnwidth,clip]{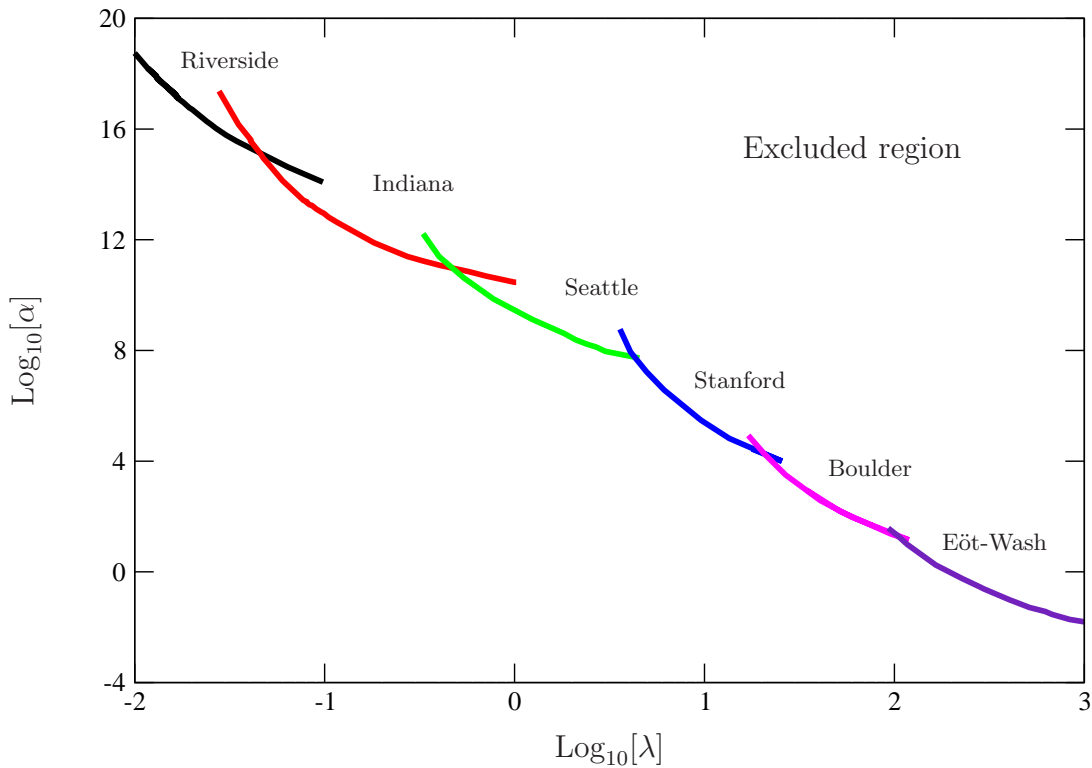}
\caption{Experimental constraints and predictions for a hypothetical 
Yukawa force in the $\alpha-\lambda$ plane, with $\lambda$ expressed in 
micrometres. The plot may be considered a prolongation of the $\alpha-\lambda$
plot of Fig. 1 on the upper left corner. The excluded region is delimited by curves
compatible with no Yukawian-signal in the measured data within two
standard deviations. From left to right, limits coming from three 
experiments on Casimir forces performed in the 100 nm-1 $\mu$m range 
at room temperature \cite{Mohideen2,Lamoreaux,Decca3}, two experiments with 
microresonators operating in the 100 $\mu$m distance range \cite{Chiaverini,Long}, 
and one experiment using a torsion pendulum with rotating attractor
\cite{Hoyle}, are shown.}
\end{center}
\end{figure}

\section{Precision and accuracy in Casimir force experiments}

To gain more insight into the reliability of the limits on
non-Newtonian forces, we briefly review some elementary notions 
of data analysis \cite{Bevington}. The {\sl precision} of an
experiment is commonly defined as the measure of how exactly the 
result is determined, regardless to what that result means in the 
framework of a specific model or theory. 
Any experimental set-up has a source of intrinsic uncertainties 
leading to different results for the repeated measurement of the 
same observable and the precision is a measure of how reproducible 
the result is in {\sl itself}. 
If the random errors result from instrumental uncertainties, 
the precision is increased by using more precise instruments. 
If instead the random errors result from statistical fluctuations of 
counting finite sequences of events, the precision will be increased 
by counting more events. In an optimized experiment, both instrumental
and statistical errors are minimized and made approximately 
equal to each other. The precision of the experiments, in the case of 
apparatuses aimed at measuring Casimir forces, should be expressed in
terms of the spreading of the result in the observable directly
measured, which can be the compensating voltage in the closed-loop actuation scheme 
for the Lamoreaux's torsion balance, the electric signal from 
the quadrant photodiode for the Mohideen configuration, the 
bridge unbalance voltage in the Bell Labs measurement, or 
the frequency shift in the Padova and Indiana measurement schemes.
These primary observables are then indirectly translated into 
a measurement of force through the physics of the detection 
system and a proper calibration with a controllable force. 
At this point, although the force is always an indirectly 
measured quantity, it may be convenient, for the sake of 
standardization among different measurement techniques, to 
express the precision in terms of error bars for the force measurement. 

The {\sl accuracy} of an experiment is instead a measure of how close the 
result comes to the true value, and then is a measure of 
how much a test of a physical theory is stringent. 
The accuracy of an experiment is therefore dependent on how well 
the experimenter can skillfully control, compensate for or take into 
account in the data analysis for systematic errors. 
Improvement of the accuracy requires first of all high precision 
data, since a given targeted accuracy needs a precision 
at least as good. However, on the top of the precise data, one 
also needs to control sources of discrepancy with respect to 
a theory to be tested. Therefore a targeted accuracy relies on precise 
data plus auxiliary information about the modelization of the 
experiment in a given theoretical framework. In other words, 
while the precision of a measurement is a concept intrinsic to 
the apparatus and the protocol used to take data, the accuracy 
is a model-dependent concept and gets better and better as we 
refine the factors which take into account the sources for 
the so-called systematic errors. Once all known factors 
determining the accuracy are well assessed, a systematic 
error left over may be the manifestation of new physics. 

It is evident that this logical procedure has to be taken with 
great care before making strong claims on new physics, and all 
possible conceivable sources of discrepancy must be scrutinized prior 
to a positive claim. All fields of physics (especially in recent 
times mostly due to sociological reasons \cite{Sociol}) have examples 
of claims for evidence of new physics which were then withdrawn after 
closer scrutiny of the data and/or the running of independent
experiments in similar conditions and under less hasty circumstances. 
These considerations, when applied to the experiments measuring 
the Casimir force, are translated into a long list of systematics 
which may mimic new effects. With respect to the bare Casimir 
formulae obtained in idealized geometries for perfect conductors 
electrically neutral at zero temperature and in ideal vacuum, there 
are indeed many possible systematic effects to be taken into account. 

\vspace{4.0mm}
\noindent
{\sl Non-ideal geometries}

There are several deviations from ideal geometries, such as 
ideal spheres and infinite parallel plates. In practice, all 
surfaces manifest a finite roughness, have finite size 
which always implies consideration of border effects, finite 
parallelism, and the geometrical parametres are determined 
with finite precision. Here is a list of considerations:

\vspace{2.0mm}
\noindent
a) Roughness of the surface. The exact formula is derived, whenever a 
first principle derivation is available, assuming that there is ideal 
reflection on the walls of the cavity. Any realistic surface will have 
roughness and therefore the scattering of the photons will be more
adequately schematized in terms of diffusion, rather than pure reflection;

\vspace{2.0mm}
\noindent
b) Finite size of the test objects and border effects. Planar surfaces
of finite size are always considered in realistic experiments. 
Therefore the predictions for infinite planes have to be corrected. 
This is particularly important in the parallel plane configuration as the 
{\sl quantum field lines}, using an electrostatic analogue, are
expected to manifest deviations from the symmetrical geometry. 
This effect may become relevant at large distances between the 
test bodies. Recently, Gies and collaborators have numerically 
evaluated border effects for a scalar field in the parallel plane
configuration. In this case, one may introduce an effective surface
area $S_{\mathrm{eff}}$  related to the geometrical area of the flat
surfaces $S$, its perimeter $C$, as the distance between the plates
$d$ as $S_{\mathrm{eff}}=S+0.36 C d$, with a correction to the 
Casimir force linear, at leading order, in the distance $d$ 
\cite{Giesprivate,Giesnew};

\vspace{2.0mm}
\noindent
c) Finite parallelism whenever applicable. The deviation 
from ideal parallelism, for instance in the case of 
a parallel plane geometry, must be kept under control to 
aim at high accuracy. The effect is relatively more important 
at small distances between the test bodies. 
The correction due to finite parallelism in the proximity 
force approximation has been discussed within the PFA scheme 
in \cite{Mosteparall};  

\vspace{2.0mm}
\noindent
d) Measurement of the size of the test objects, for instance the 
radius of the sphere, the size of the plates, and the distance between
the two objects. The use of electron scanning microscopy allows one to 
obtain precise measurements of some geometric parameters, for instance
the diameter of the spheres used in the AFM-MEMS experiments by 
the Riverside, Bell Labs, and Indiana groups. One critical issue 
still awaiting a definitive solution is the precise 
determination of the distance between the two objects. 
Indeed, the {\sl relative} distance between two different 
configurations corresponding to different distances can be 
accurately obtained by calibrating the displacement actuators, 
but the {\sl absolute} distance is not, and this is important 
for forces strongly dependent on the distance. 
While in experiments using dielectric surfaces this problem 
was addressed by using optical interferometric techniques, in various
experiments performed so far with metallic surfaces the 
absolute distance was determined as a fitting parameter. 
When using fiber optic interferometry, two-color fiber 
optic displacement transducers  like those of the Indiana group 
will be a standard  solution for determining the absolute distance 
with an accuracy comparable to the determination of the two wavelengths.
Other solutions could include the use of calibrated wires 
of small diameters (available down to 2-5 $\mu$m) as high precision 
spacers to assess for few experimental points the absolute distance. 
All these solutions have limitations. 
The more conceptual one is that, at the atomic level, the distance between 
two bodies is an ill-posed concept as surfaces are far from being 
accurately schematized as planes.  It is then hard to assess the
distance with a precision better than the surface roughness evaluated 
as the rms deviation over the considered surfaces. The same definition 
of distance may depend also on the probe used to assess the distance. 
In case of electrostatic calibrations, non locality in the electric 
properties of the surfaces may be a source of systematic errors. 
In the case of the use of fiber optic interferometry, light is not just reflected
by the first atomic layer, and the procedure of obtaining the zero 
distance reference by touching the surfaces has to be taken with great
care, due to potential uncertainties inherent to this somewhat subjective, not 
necessarily reproducible, procedure. 

\vspace{4.0mm}

\noindent
{\sl Electrical properties}

Ideally one needs materials which are perfect reflectors for 
photons at any wavelength, and which are electrically neutral.
None of these conditions is obtained in practice: 
 
\vspace{2.0mm}
\noindent
a) There is no perfect conductor at all frequencies, so at frequencies 
large enough the photons will leak out from the cavity and will 
not contribute to the radiation pressure;

\vspace{2.0mm}
\noindent
b) The presence of residual electric charges will spoil the 
neutrality and will give a direct Coulomb contribution. We can
identify at least two distinct sources of residual electric charges, 
respectively due to the so-called patch effects, and the Volta 
potentials. The former are intrinsic to the same metal, while the 
latter refers to the differences in the mean electric potential between two 
connected metals.

The physical origin of the surface potential and charge on a metal 
is due to the presence of a finite surface dipole moment. This depends 
on the separation of the lattice planes parallel to the surface. 
If there are different crystallographic directions on the surface of a
polycrystalline metal, there will be variations of the surface potential. 
The effect of such forces in the Casimir experiments has been 
discussed in detail, using analytical and numerical methods, in 
\cite{Speake}. The mapping of the surface voltage can be obtained 
by using Kelvin probes (see \cite{Schroder} for a review), and 
measurements on various conductors, including gold, copper and 
aluminum, have been performed with a spatial and potential 
resolution of 1.5 mm and 1 mV, respectively \cite{Camp1}. 
The same authors have tried to contain the surface potentials 
by focusing on treatment of nonreactive materials like gold and 
graphite \cite{Camp2}. In particular, sublimation of fullerene films 
on metallic substrates has shown that there are no surface potential 
variations within the Kelvin probe resolution of 1 mV \cite{Camp3}. 

The presence of contacts between different metals allows also 
for different mean voltages. For instance, in the Lamoreaux 
experiment the two  plates were externally shorted through a 
loop of 40 separate electrical connections, leading to an 
overall Volta potential of 430 mV \cite{Lamoreaux}. 
This results in an electrostatic force which overwhelmed the 
Casimir force. Even more important, the combination of this electrostatic 
attraction between the two plates and the finite retaining force 
of all experimental configurations results in stiction of the surfaces 
at relatively large distances, thus preventing the achievement of the 
smaller gaps at which the Casimir force is expected to prevail over the 
electrostatic contribution. Provided that the drift in time of the
effective Volta potential is minimal, the solution adopted consists in
the application of a {\sl counterbias} voltage cancelling, at least at 
the leading order, the Volta potential contribution. This allows for
the achievement of smaller gaps to the point where the stiction will 
result from the Casimir force itself. The residual voltage difference due to the 
incomplete cancellation can be taken into account in the off-line 
data analysis through a simultaneous fitting with the expected 
Casimir and Coulomb forces. 

\vspace{4.0mm}

\noindent
{\sl Environmental properties}

The measurement of the Casimir force is also affected by environmental 
factors. The leading ones are due to the real photons present in the 
blackbody radiation at finite temperature, the presence of residual 
gas in the gap, and the acoustic noise in the laboratory. So far, 
all experiments have been performed in conditions such that magnetic 
fields should have no effect, although at the microscopic level this 
requires a control over magnetic properties of the test masses or 
magnetic impurities present in them which eventually could be relevant 
for accurate measurements. 

\vspace{2.0mm}
\noindent
a) Finite temperature. An obvious effect of finite temperature, at 
the instrumental level, is the presence of thermal drifts in 
the apparatus, which may limit the effectiveness of taking long 
integration times. The use of Peltier coolers with active feedback 
and materials with low expansion coefficients mitigates this effect. 
Moreover, at any finite temperature one also expects a blackbody 
background of photons. Most of these photons are blocked by the cavity 
as the condition of stationarity for the electromagnetic field implies 
that no electromagnetic waves with wavelength exceeding twice the
length of the cavity can be sustained. The thermal correction for a
given temperature is particularly important at large distances.
This effect is included in the realistic approach pioneered by
Lifshitz, but complications arise due to the necessary inclusion of  
the reflectivity coefficients. 
In the Lifshitz formula for the force between two parallel surfaces \cite{Lifshitz}:

\begin{equation}
F_C= \frac{S}{\pi \beta d^3} \sum_{m=0}^{\infty \;  
'} \int_{m \gamma}^{\infty} dy \; y^2
\; \left[  \frac{r_{\rm TM}^{-2} e^{-2y}}{1-r_{\rm TM}^{-2} e^{-2y}}  +
\frac{r_{\rm TE}^{-2} e^{-2y}}{1-r_{\rm TE}^{-2} e^{-2y}} \right]
\label{forcepp}
\end{equation}
both finite temperature (with $\beta=1/k_{\rm B} T$ the inverse
temperature, and $\gamma=2 \pi d / \beta \hbar c$), and finite
conductivity effects (expressed through the reflection coefficients 
$r_{\rm TE}$ and $r_{\rm TM}$ for the two independent polarizations 
TE and TM computed at imaginary frequencies $\omega_m=i \xi_m$, 
where $\xi_m=2 \pi m / \beta \hbar$ are the Matsubara frequencies) 
are taken into account. The $m=0$ term is counted with half weight, as 
denoted by the prime on the summation sign. 

The reflection coefficients are expressed in terms of the dielectric 
permittivity $\epsilon(\omega)$ as
\begin{eqnarray}
r_{\rm TM}^{-2} = 
\left[ \frac{\epsilon(i \xi_m) p_m + s_m}
{\epsilon(i \xi_m) p_m - s_m} \right]^2  &;&
r_{\rm TE}^{-2} =\left[ \frac{s_m + p_m}{s_m - p_m} \right]^2 ,
\end{eqnarray}
where $p_m=y/ m \gamma$ and $s_m = \sqrt{\epsilon(i  \xi_m) - 1 + p_m^2}$.
Using tabulated optical data for different metals in the bulk, it is 
possible to compute the dielectric permittivity along the imaginary 
frequency axis. To calculate the $m=0$ contribution, the available 
data have to be extrapolated in the static limit. 
There is no unique prescription to do this.  Different choices have led to
controversial predictions for the Casimir force between parallel 
plates (see for instance \cite{t1,t2,t3,t4,t5,t6,t7,t8,t9,t10,t11,t12,Miltonrecent}
and the in-depth discussion in \cite{Milton1}).
The definitive control over the thermal fluctuations will be required to 
give limits to other forces. Detailed data analysis such as the one
proposed in \cite{Decca2} cannot be used to simultaneously constrain
two physically different effects as the thermal contribution and the 
presence of gravitational forces of Yukawian origin in absence of 
other experimental inputs;

\vspace{2.0mm}
\noindent
b) Residual gases can influence the measurement both by giving 
additional random forces acting on the test bodies, by  
changing in time the properties of the surfaces, and by  
favoring the migration of contaminants, oxide layers, dust particles, 
and residual charges; 

\vspace{2.0mm}
\noindent
c) Acoustic and seismic noise. The experiments should be shielded 
from mechanical noise originating either by natural sources, as 
seismic noise, for instance using air tables and damping systems, and 
from artificial sources including vacuum pumps. 
In some cases, this can limit the measurement time as the vacuum chamber 
must be disconnected from the vacuum line, as described in
\cite{Decca4}. In the Padova experiment, the demand for vacuum 
also comes from the use of a scanning electron microscope evacuated 
through a diffusion pump, which was in turn periodically fed by 
a (quite noisy) roughing pump.  In the case of the extremely sensitive 
torsional balance set-up, Lamoreaux reported that his own presence 
in the laboratory during the measurement, and the consequent tilting 
of the floor due to the additional weight in the room, was a factor 
initially affecting the data \cite{Lamoreauxhisto}. 

\vspace{2.0mm}
Furthermore, even for idealized physical and geometrical conditions,
one often has to deal with the proximity force approximation formula 
in the case of sphere-plane or cylinder-plane geometries, and with 
data analysis where often unambiguous choices cannot be made. 
The use of the proximity force approximation has generated 
a wide debate, since it is used in the sphere-plane configuration for 
which an exact formula does not exist and, due to this reason, 
it is then hard to assess its validity. Under idealized 
conditions for all the other possible sources of systematics, 
it is generally believed that the PFA is valid within 0.1 $\%$ 
for large radii of curvature \cite{Bordag1}. Recently, progress 
has been made by various groups towards understanding the limitations 
of the PFA and its interplay with some of the abovementioned effects. 
The Paris group led by Lambrecht and Reynaud \cite{MaiaEPL} has 
discussed the interplay between roughness corrections, assuming 
the plasma model for conductivity corrections, and the PFA, 
showing that the roughness correction is underestimated in the PFA framework. 
The Casimir energy between a cylinder and a plate has been evaluated exactly 
\cite{EmigMIT,Bordagnew}, allowing for a comparison with the 
corresponding PFA formula previously evaluated in \cite{Dalviteccentric}. 
The worldline numerics approach pioneered by  Gies and collaborators
\cite{Gies1} has been applied to both sphere-plate 
and cylinder-plate geometries for a scalar field with Dirichlet
boundary conditions, showing significant deviations from PFA if the 
ratio between the gap and the radius of curvature of the surfaces is 
too large \cite{Gies2}. 

The systematic factors listed above do not contribute with equal weights 
in the entire range of explored distances. Finite roughness, finite
parallelism, finite conductivity, vacuum pressure, and Volta potentials 
more strongly affect the measurement at small distances. 
The finite size and border effects, the finite temperature
contribution, and the proximity force approximation (whenever applied) 
are corrections expected to play a major role at larger distances. 

Another issue to be experimentally assessed is the sensitivity
to the minimum detectable force using physical measurements. 
For larger scale experiments (like torsional balances) this 
is better assessed for instance in terms of the maximum distance 
at which the gravitational force is still detectable. 
For smaller scale experiments (like AFM-based ones) the sensitivity 
should be naturally expressed in terms of the minimum detectable
electric field. There are also other physical effects to be exploited 
to assess the precision and the sensitivity of the measurements in 
the short range. In particular, the conducting properties of the 
substrate can be used, for instance using semiconducting surfaces 
with modulated conductivity. The sensitivity could then be expressed 
in terms of the minimum charged carrier modulation detectable through 
the Casimir force measurement. Even simpler, the Casimir force is
expected to depend on the thickness of the substrate due to the finite skin depth, as 
evidenced in \cite{Iannuzzi3}. The claimed accuracy of the measurement 
could be achieved with a {\sl blind} test, by providing two surfaces 
having a thickness unknown to the same Casimir force experimenter. 
If the claimed accuracy is real, he/she should be able to infer, from 
the Casimir force measurements, the difference between the thickness 
with a corresponding confidence level, to be compared with actual 
measurements of the thickness of the substrates performed by a third
party. This unbiased benchmark could be the ultimate test of the 
reliability of the accuracy claimed in each experiment. 
Similar benchmarks are also available by detecting the effect of
external fields, like the presence of a magnetic field \cite{Wang}, 
in a controllable way. In principle, also effects like the so-called 
{\sl lateral} Casimir force \cite{Mohilater} could help to assess 
the sensitivity, however the control over the related theoretical 
predictions based upon the PFA seems to be an issue \cite{Rodriguez}. 
 
A further potential concern for the assessment of
the limits to non-Newtonian forces also arises from the use of 
electrostatic calibrations at larger distances. 
For instance, in all the experiments determining the absolute distance 
through a fitting to the purely electrostatic data, one cannot
logically expect strong and meaningful limits to non-Newtonian forces. 
In doing this, one first insulates a region of parameters, at a large 
distance where a purely electrostatic contribution is expected.
In this regime, all groups do not consider any of the usual
corrections included in the evaluation of the Casimir force, something 
which may limit the accuracy of the calibrations in the first place.  
For some corrections (like for instance roughness) this is partially 
justified since the electrostatic calibrations are performed at such 
large distances that they are negligible.   
From this analysis in a partial region one extract some parameters, 
for instance the absolute distance and the Volta potential, and then 
this determination is used in a region where the Casimir force is 
expected to give a significant contribution. 
This works fine for the latter force due to the sharp dependence on 
distance, but the expected Yukawian contribution may fall off slowly at 
large distance depending on the value of the Yukawa range $\lambda$, 
so to first neglect this term in the fitting of the electrostatic 
calibrations and then to look at it in the residuals of the Casimir 
force has a somewhat tautological flavor. 
In particular, a Yukawian force of relatively large intensity and
long range could be hidden in the electrostatic calibration. 
Its presence would be evidenced by a less accurate fit of the data 
in the long distance with the Coulomb contribution alone. 
However if the fitting parameters are then used for obtaining
information on the Casimir force it will be very hard to infer 
from the residuals the presence of this Yukawian force. 
To avoid such loopholes, a combination of improvements should be 
implemented, first of all a precise determination of the distance 
between the test objects independent on the electrostatic calibrations, 
and thereafter a fitting procedure in the widest range of 
distances with the Casimir force, the residual electrostatic force, 
and the hypothetical Yukawian force all considered at the same time. 

As should be clear from previous discussions, it is very important to 
analyze the data for extracting information about the windows of
opportunity for non-Newtonian forces. A possible approach to data 
analysis which should help comparison of various experiments and
assessments on non-Newtonian limits is based on the following argument. 
As remarked ealier, Casimir physics is intrinsically ruled, at the 
leading order, by the presence of the two fundamental constants 
present in relativistic quantum field theory, namely $\hbar$ and $c$. 
In all formulae for the Casimir effects the two constants appear 
in a multiplicative way. All the experiments on Casimir forces 
can then be rephrased as macroscopic determinations of the 
product $\hbar c$.  This has the advantage that it allows for 
a direct comparison of experiments performed in a variety of geometries. 
Moreover, since high precision measurements of $\hbar$ 
($c$ being now exactly determined) are independently available, 
deviations of $\hbar c$ obtained from Casimir experiments with respect 
to the more precisely known value of $\hbar c$ allow for a unified 
discussion of the accuracy of each experiment. The accuracy of the 
measurement can be expressed in terms of the relative difference 
between the  central value for $\hbar c$ obtained from the data 
fitting of the Casimir measurements and its {\sl true} value 
({\it i.e.} with much smaller error) as available from CODATA, 
$\hbar c= (3.16152636 \pm 0.00000054) \times 10^{-26}$ J m.   
Deviations of statistical significance from the CODATA value 
obtained from the data analysis of the experiments on Casimir force 
should signal either a not well controlled background or new 
physics in the micrometre range. The latter hypothesis could become 
stronger if confirmed by similar pattern of deviations in independent 
and quite different experimental set-ups.

\section{Future experimental directions}

In this section we summarize some directions currently pursued to 
understand the long-distance behaviour of the Casimir force 
which are relevant for limits to non-Newtonian forces. 
The current status and future directions of short-distance Casimir physics, 
for instance the interplay of quantum fluctuations with surface collective 
excitations \cite{Intra1,Intra2}, will be more adequately discussed 
in other review papers of this Focus issue.

Among the directions to be explored, there is the possibility of
improving the AFM sensitivity to detect the Casimir force in the 
1-2 $\mu$m range or to use differential techniques to measure the 
force at smaller distances at different temperatures \cite{66}, as 
proposed by Mohideen and collaborators.  At large distances, the 
$1/d^3$ scaling of the sphere-plane Casimir force is more favourable 
than the $1/d^4$ scaling of the force between parallel plates, and the 
expected thermal corrections are more pronounced \cite{Dalviteccentric}. 
In particular, high precision AFM apparatuses can be developed to reach sensitivity in the 
$10^{-18}$N/$\sqrt{\mathrm{Hz}}$ range by using cryogenic techniques. 
Apart from the broader technological benefits of developing high precision 
AFM apparatuses, this direction can lead to better limits on
non-Newtonian forces without the cross-talk with the still not 
completely understood thermal correction as lower temperatures 
correspond to longer thermal effective lengths. 
Pursuing this direction will require a well focused effort with 
the need to adapt a Casimir force apparatus in a cryogenic environment. 

\begin{figure}[t]
\begin{center}
\includegraphics[width=0.7\columnwidth,clip]{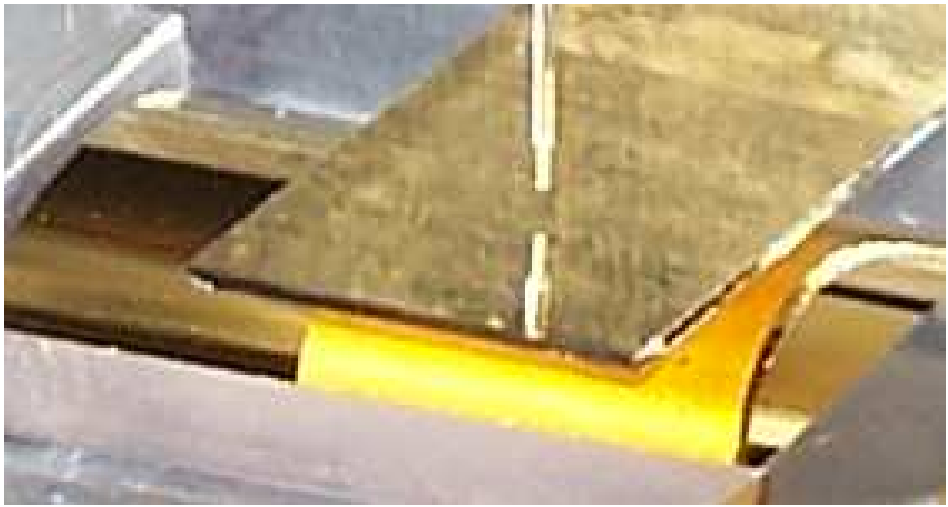}
\caption{Image of the experimental set-up for the study of 
the cylinder-plane configuration, with (from top to bottom) the 
optical fiber for the displacement measurement, the silicon rectangular 
cantilever, and the cylindrical lens. Cantilever and cylindrical 
lens, coated with (144$\pm$ 20) nm of Au, have respectively width and radius 
of curvature equal to 1 cm. Notice the mirror image of the optical 
fiber on the cantilever, which allows to estimate the fiber to 
cantilever distance. With proper illumination and the use of an optical 
microscope by looking at the mirror image of the cylinder on 
the cantilever, it is likewise possible to estimate the cylinder-cantilever 
distance at large (above $\simeq 10 ~ \mu$m) gaps, which allows for 
consistency checks with the electrostatic calibrations.}
\end{center}
\end{figure}

A second direction to be pursued is an improved version of the 
experiment using a torsional balance. This direction is under 
development both at Los Alamos for the sphere-plane geometry
\cite{Lamotor,Buttler} and in Grenoble for the parallel plate 
geometry \cite{LambrechtCQG}.
Lamoreaux has discussed a series of improvements to his original
set-up. The continuation of his efforts seems quite natural and this 
could clarify open issues related to the original measurements reported in 1997. 
Meenwhile, a torsion balance for a measurement of the Casimir 
force between large surface parallel plates has been built in at ILL 
in Grenoble and electrostatic calibrations are currently being performed.  
The experiment will also make use of already developed inclinometer
technology used to measure the quantum states of neutrons in the 
gravitational field \cite{Nesvy1} to deal with issues of
parallelization and absolute measurement of the gap distance.
Even in this case an improvement of the sensitivity is expected, at 
the price of a longer development stage, using cryogenic techniques 
\cite{Bantel}.

A third direction actively pursued is the study of the Casimir force 
in a cylinder-plane geometry. This configuration has advantages and 
disadvantages in between the sphere-plane and the parallel plates geometries. 
The Casimir force in the cylinder-plane configuration, evaluated in the 
proximity force approximation, is 
\cite{Dalviteccentric}

\begin{equation}
F_C(d) = \frac{\pi^3 \hbar c}{384 \sqrt{2}} \frac{L a^{1/2}}{d^{7/2}},
\end{equation}
where $L$ is the length of the cylinder, and $a$ its radius of 
curvature. This configuration looks particularly promising to 
detect the thermal effect and allows for an easy procedure to 
obtain accurate parallelization. An exact solution for the 
Casimir force has been found recently \cite{EmigMIT}, and 
thus this configuration is also a candidate to test 
the validity of the proximity force approximation in case the 
cylinder is made of small (around few $\mu$m diameter) wires.
An experiment in this configuration is now under development at Dartmouth with 
electrostatic calibrations having shown that the measurement should 
be feasible \cite{BrownPRA}. As of this writing the first data acquisition 
with an upgraded apparatus aimed at measuring the Casimir force in 
the 2-5 $\mu$m range is underway (see Fig. 4).  

\section{Conclusions}

\vspace{0.3cm}
\noindent
We have reviewed past experiments which have successfully measured the
Casimir force, and from which limits on the existence of
non-Newtonian forces have been derived. 
We have also discussed some of the future directions which are 
being explored with the aim of better mastering the systematics 
of the corrections to the Casimir force, in particular the urgent 
issue of the finite temperature correction. Without a solid 
understanding of this effect, limits to non-Newtonian forces 
appear very weak and prone to criticism. This in turn demands 
reliable assessments of the precision and sensitivity of the 
experimental set-up, for instance also by means of blind 
tests as proposed in Section IV. 
It will be also necessary, in order to avoid unintentional bias in 
the data analysis and to collect diverse feedbacks, to allow for 
a broad {\sl data sharing} among various experimental groups and 
phenomenologists interested in independent analysis. 
This is possible at zero cost through internet resources and simply
adopting what is already implemented in other subfields of physics, most
notably in astronomy and high-energy physics, with data archives 
available to all interested astronomers and physicists. 

All the demonstrations of the Casimir force performed so far 
have not provided evidence, within their claimed accuracy, for 
any residual signaling non-Newtonian physics. 
We have briefly discussed a proposal to parametrize the data 
collected in the six Casimir force experiments, or future experiments,
in terms of a measurement of the constant $\hbar c$. 
Due to the long list of systematic effects, it is also clear 
that an extremely careful and conservative analysis has to be 
carried out prior to any claim of new physics. This could be 
evidenced as a significant deviation between the central value 
of the $\hbar c$ value determined from the Casimir data and 
the $\hbar c$ CODATA value. The repetition of experiments in 
dissimilar conditions and aimed at observing the same signal 
will help disentangle a possible positive claim from false signals. 
We believe however that a definitive confirmation of a claimed effect 
will most likely happen by collecting complementary evidences 
based upon radically different classes of experiments from the one 
described here and operating at different lengthscales. For instance, 
relevant informations could also be acquired by exploiting techniques 
close to the ones developed for detecting light neutral particles 
coupled to photons by Zavattini {\it et al.} \cite{Zavattini}, 
by accurate measurements of neutron's states in the Earth's gravitational 
field \cite{Nesvy2}, by 
observing surface excitations in superfluid helium \cite{Roche,Kirichek}, 
by measuring the dynamics of intrinsically nanotechnological
structures like nanotubes 
\cite{Blagov}, or significantly improving the sensitivity of 
existing dynamical detection schemes for rotating \cite{Adelb} 
or vibrating masses \cite{Schiller1,Schiller2}.

The message of this review in a nutshell is that, in spite of the recent
progress which brought solid demonstrations of the Casimir force, the 
challenge of discovering new gravitational-like phenomena in the
micrometre range with table-top experiments is still an issue more open than ever.
A third generation of Casimir force experiments combining creativity, hard work, 
metrological accuracy, and substantial support both in intensity and duration 
will be necessary to further pursue non-Newtonian gravitation at the microscale. 
 
\acknowledgments

I acknowledge pleasant and fruitful discussions with many colleagues, along 
many years of joint efforts to understand quantum vacuum at the macroscopic 
level, in particular with J.F. Babb, G. Bressi, I. Brevik, F. Capasso,
G. Carugno, D.A.R. Dalvit, R. Decca, G. V. Dunne, E. Fischbach, L. Ford, H. Gies, 
N. Graham, D. Iannuzzi, F. Intravaia, M.-T. Jaekel, R.L. Jaffe, M. Kardar, G.L. Klimchitskaya, 
A. Lambrecht, S.K. Lamoreaux, F.C. Lombardo, F.D. Mazzitelli, P.W. Milonni, 
K.A. Milton, U. Mohideen, V.M. Mostepanenko, S. Nussinov,
L. Pitaevskii, E. Polacco, S. Reynaud, G. Ruoso, V. Svetovoy, C. Villareal and E. Zavattini. 
I am grateful to F. Capasso, R. Decca, T. Ederth, U. Mohideen and  
S. K. Lamoreaux for providing pictures of their experimental systems reported in Figure 2,  
and to M. Brown-Hayes, W.-J. Kim, K.A. Milton, and G.A. Wegner for a critical 
reading of the manuscript and many constructive comments.

\end{document}